\documentclass[%
 rsi,
 amsmath,amssymb,
 reprint,%
]{revtex4-1}

\usepackage{graphicx}
\usepackage{dcolumn}
\usepackage{bm}
\usepackage{xcolor}
\usepackage[utf8]{inputenc}
\usepackage[T1]{fontenc}
\usepackage{mathptmx}
\usepackage{etoolbox}

\makeatletter
\def\@email#1#2{%
 \endgroup
 \patchcmd{\titleblock@produce}
  {\frontmatter@RRAPformat}
  {\frontmatter@RRAPformat{\produce@RRAP{*#1\href{mailto:#2}{#2}}}\frontmatter@RRAPformat}
  {}{}
}%
\makeatother
\begin{document}

\preprint{AIP/123-QED}

\title[Sample title]{Nanoelectrospray ionization coupled to a linear charge detection array ion trap spectrometer for single viral particles analysis}
\author{S. Maclot}
\altaffiliation[Also at ]{Normandie Univ., ENSICAEN, UNICAEN, CEA, CNRS, CIMAP, 14000 Caen, France.}
\affiliation{ 
Institut Lumi\`ere Mati\`ere, University of Lyon, Universit\'e Claude Bernard Lyon 1, CNRS, Lyon
}%
\author{T. Reinert}%
\author{L. Duplantier}
\affiliation{ 
Institut Lumi\`ere Mati\`ere, University of Lyon, Universit\'e Claude Bernard Lyon 1, CNRS, Lyon
}%
\author{L. Thiede}
\affiliation{CCSSB Centre for Structural Systems Biology, Deutsches Elektronen-Synchrotron DESY, Leibniz Institute of Virology (LIV), University of L\"ubeck, Notkestra{\ss}e 85, 22607 Hamburg, Germany \&
Institute of Chemistry and Metabolomics, University of L\"ubeck, Ratzeburger Allee 160, 23562 Lübeck, Germany}
\author{G. Montagne}
\author{C. Clavier}
\author{X. Dagany}
\author{C. Comby-Zerbino}
\author{M. Kerleroux}
\affiliation{ 
Institut Lumi\`ere Mati\`ere, University of Lyon, Universit\'e Claude Bernard Lyon 1, CNRS, Lyon
}%
\author{R. Pogan}
\author{C. Uetrecht}
\affiliation{CCSSB Centre for Structural Systems Biology, Deutsches Elektronen-Synchrotron DESY, Leibniz Institute of Virology (LIV), University of L\"ubeck, Notkestra{\ss}e 85, 22607 Hamburg, Germany \&
Institute of Chemistry and Metabolomics, University of L\"ubeck, Ratzeburger Allee 160, 23562 Lübeck, Germany}
\author{A.N. Kozhinov}
\author{K.O. Nagornov}
\author{Y.O. Tsybin}
\affiliation{Spectroswiss, 1015, Lausanne, Switzerland}
\author{D. Papanastasiou}
\affiliation{Fasmatech Science \& Technology, Lefkippos Tech. Park, NCSR Demokritos, 15341 Agia Paraskevi, Greece}
\author{R. Antoine}
    \email{rodolphe.antoine@univ-lyon1.fr}
\affiliation{ 
Institut Lumi\`ere Mati\`ere, University of Lyon, Universit\'e Claude Bernard Lyon 1, CNRS, Lyon
}%

\date{\today}

\begin{abstract}
This work presents the implementation of a new charge detection mass spectrometer (CDMS) design that operates in a stand-alone mode, thanks to its integration with nanoelectrospray ionization. More specifically, this innovative CDMS consists of a linear charge detection array ion trap spectrometer that combines an eight-tube detector array with conical electrodes. This configuration allows for recording data in both transmission mode (linear array) and ion trapping mode (ConeArrayTrap), which enables the measurement of time-of-flight (related to the mass-to-charge ratio) along with the charge of individual ions. As a result, this design supports high-throughput metrology of viruses at the single-particle level. The devices and geometry of the instrument have been developed based on ion optics simulations. The performance of the current instrument is demonstrated using human norovirus-like particles (hNoVLP) and Adenovirus Ad(5) (hAdV5).
\end{abstract}

\maketitle

\section{Introduction}
The coronavirus' extraordinary ability to spread in the environment has established the SARS-CoV-2 pandemic as one of the greatest challenge facing humanity in the 21st century.\cite{zoumpourlis2020COVID}
Protective measures based on monitoring coronavirus dispersal in the environment and rapid characterization of individuals have become essential to ensure the safety of our ageing population and to restart/sustain the global economy.\cite{maqbool2024COVID} Mass spectrometry (MS) is a very commonly used technique with extraordinary sensitivity, and over the last 20 years virology found mass spectrometry to be a key tool.\cite{milewska2020MS,duelfer2019MS}
MS is crucial for identifying viral particle content, aiding studies on viral protein composition and vaccine development.\cite{lewis1998MS}

In this work, we present the nanoelectrospray ionization source platform coupled to linear charge detection array ion trap mass analyzer setup, which is part of the ARIADNE-Vibe platform operated in stand-alone mode  at Lyon University.\cite{ARIADNE} The technical objective of the ARIADNE-Vibe project is to develop an analytical platform based on multidimensional MS instrumentation.With the aim of performing direct and instantaneous detection of intact viral particles in breath and water, this unique analytical platform will push the boundaries in all aspects of MS-centric analytical sciences, incorporating a series of disruptive technologies integrated into a single system.

Currently, viruses, bacteriophages, and small infectious agents challenge MS measurement capabilities, yet the growing precision on MS instruments supports its use in viral analysis. Key MS-based techniques for high molecular weight detection include charge-detection mass spectrometry (CDMS) \cite{keifer2017CDMS,antoine2020CDMS,D1MA00261A} and differential mobility spectrometry ‐ based analyzers like gas‐phase electrophoretic mobility molecular analysis \cite{kaufman1996GEMMA,weiss2019GEMMA}.
Fuerstenau et al. were the first to measure masses of intact viruses by CDMS.\cite{fuerstenau2001MSvirus}
These measurements were performed by single-pass CDMS. Obtained mass spectra on rice yellow mottle virus (RYMV) and tobacco mosaic virus (TMV) presented low resolution and limit of detection was ~ 450 elementary charge (e). Antoine et al. further used TMV virus, with characteristic rod-like structures, as calibrant for amyloid fibrils mass spectra.\cite{doussineau2016amyloid}
Jarrold’s group pushed the limits of CDMS for determining stoichiometry of protein complexes, viruses, and virus assembly. \cite{keifer2017CDMS,jarrold2024CDMS,jarrold2021CDMS}

\begin{figure*}
 \includegraphics[width=0.9\textwidth]{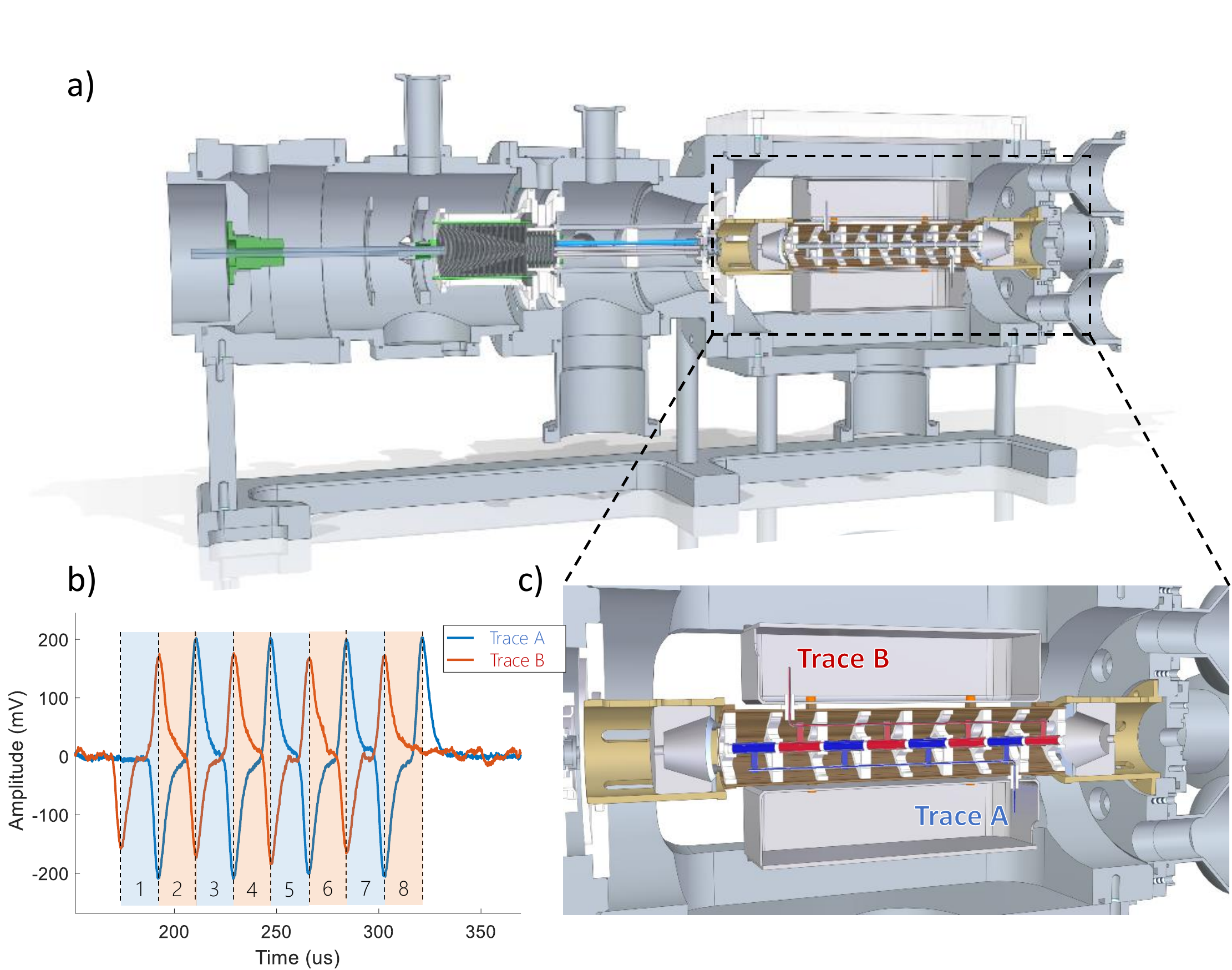}
 \caption{\label{fig:apparatus} a) General view of the apparatus. b) Example of signal traces (droplets) coming from the 2 sets of 4 CDDs clearly visible in the zoom-in panel c). }
\end{figure*}

 The aim of this work is to provide a cutting-edge CDMS platform for high throughput and high resolution measurements of viral particles performed in a single particle counting mode. CDMS measurements have been realized in three different ways, each having unique strengths and weaknesses.\cite{https://doi.org/10.1002/mas.21920} The easiest but the least accurate mode of operation is single-pass CDMS where each ion travels through the analyzer only once. \cite{doussineau2011CDMS_PEG} Nevertheless,  1000s of ions/s can be detected, and averaged data to improve accuracy can be collected relatively fast.The more accurate approach exhibiting far lower limit of detection involves operating a CDMS analyzer in trapping mode. \cite{doi:10.1021/acs.langmuir.3c02526}
 In CDMS ion traps, one or more ions are trapped between two ion mirrors and detected multiple times, thus improving the signal-to-noise ratio of the measurement. \cite{doussineau2011CDMS_trap} In this trapping mode of operation data collection is slower (1 ion/10 s) \cite{CONTINO2013153}, and it takes several hours to record mass spectra. The trade-off between speed and accuracy is balanced in a linear array CDMS where each ion travels only once through a series of detectors. \cite{gamero2007CDMS,smith2011CDMS} A linear array consists of a series of cylindrical detector electrodes arranged coaxially, for instance, eight interleaved detectors grouped in two sets of four, with each set attached to a separate amplifier, as shown in Fig.~\ref{fig:apparatus}. The main advantage of this configuration is the higher throughput (up to 100 ions/s) combined with enhanced signal amplification afforded by the multiple detector scheme. Most recent advances led by M.F. Jarrold and E.R. Williams involves the development of such charge detector arrays while operating in ion trapping mode  to further improve the sensitivity and precision of this technique. \cite{keifer2017CDMS_single,elliott2017singleCDMS}. D.E. Austin's team, meanwhile, is working on miniaturizing this technique on printed circuit boards.\cite{barney2013Boards}. A. J. R. Heck's team and N. L. Kelleher's team have developed orbitrap-based CDMS instrument and methods, that are currently commercialized by ThermoFisher supplier.\cite{desligniere2024orbitransient, WOS:000517743000002}
 
In this work we have developed a new CDMS configuration by implementing a linear detector array including two end-cap electrodes, which allow both trapping and transmission modes to be evaluated in a single platform. The linear detector array consists of a series of cylinders (8 interleaved detectors with 2 amplifiers, see Fig.~\ref{fig:apparatus}). The possibility to add ion mirrors (ConeTrap) at the ends of 8 identical charge detection devices (CDD) permit a trapping mode to significantly increase the resolution. This ensemble (coined ConeArrayTrap) is used here for multiply charged viral particles produced by nanoelectrospray ionization.

 \begin{figure*}
 \includegraphics[width=0.9\textwidth]{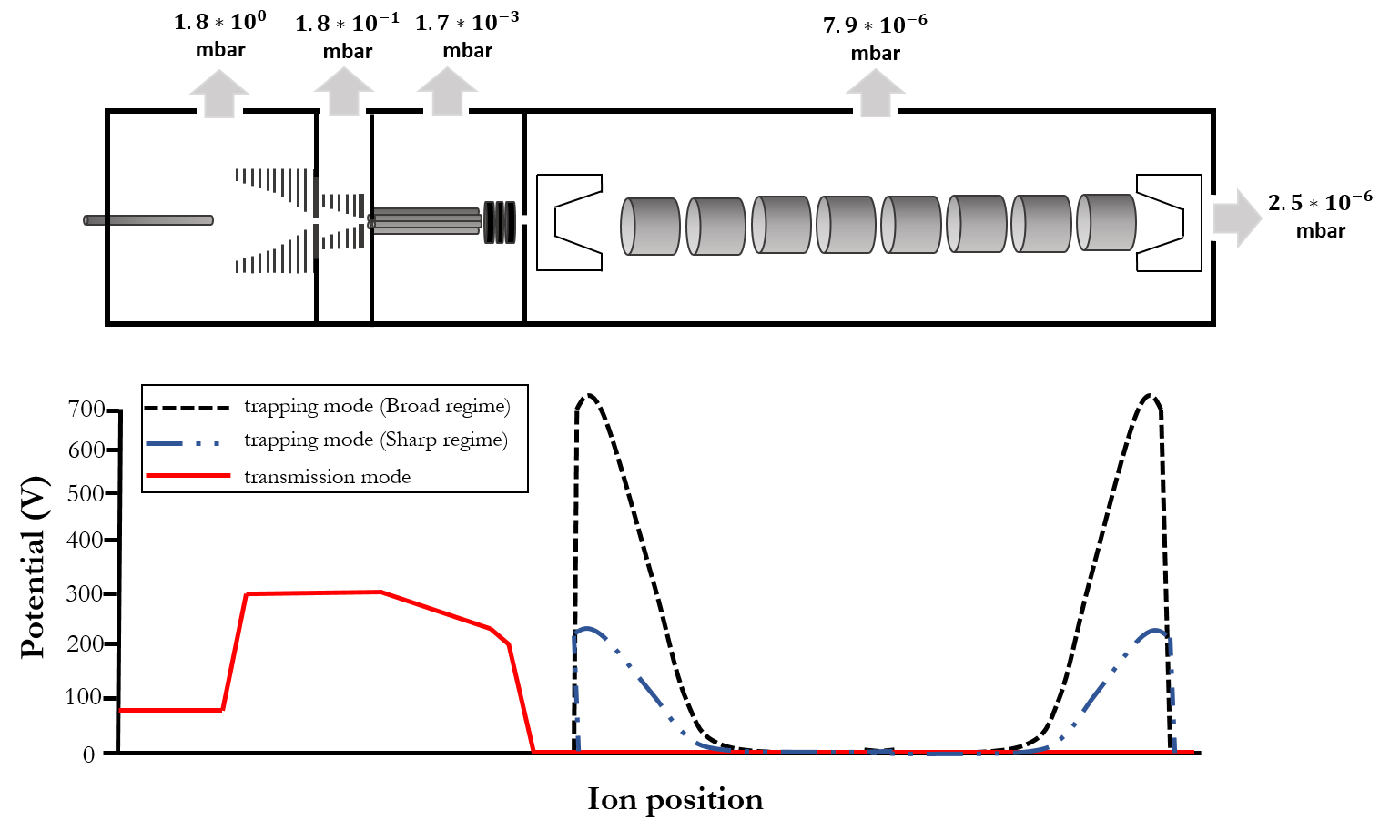}
 \caption{\label{fig:other} Scheme of the constructed CDMS instrument used in this work. The DC experienced by the ions through their position in the CDMS are presented for the transmission mode and the two trapping modes (sharp and broad). }
\end{figure*}
 
 \section{Apparatus description}
 \subsection{General description}
Figure~\ref{fig:apparatus} provides an overview of the general design of the charge detection apparatus, with an overall length of 1.5 m. The machining of mechanical parts was produced within ISO-Js12 tolerance (SMGOP Fontaine, France). Ions produced in the nanoelectrospray source (left side on Figure~\ref{fig:apparatus}) are subsequently introduced through an inlet capillary into the first vacuum stage accommodating a dual RF ion funnel operated at 742 Hz, 1182 Hz and 300 V at $1.8\times10^{0}$ and $8.6\times10^{-2}$ mbar.
Ions focused through a 2 mm differential aperture at the exit of the second ion funnel are subsequently transferred through an hexapole ion guide installed into the third vacuum stage operated at  $9.8 10^{-4}$ mbar into the CDMS analyzer. This radio-frequency (RF)-hexapole ion guide is terminated by an einzel lens controlling beam divergence transfer of ions with minimal losses through the entrance end-cap of the CDMS analyzer. The pressure in different chambers of the instrument are given Figure~\ref{fig:other}).
\subsection{Nanoelectrospray and ionic train}
The nanoelectrospray ion source is based on the Nanospray Flex Ion Source design (Thermo Scientific) which was customized \cite{D2CC00200K} to fit the entrance of the CDMS frame. The ionization source consists of a housing with manual XYZ-manipulator and fittings, a DirectJunction adaptor for online analysis and a camera set-up coupled to an LCD monitor to visualize the spray. 
Electrospray capillaries were homemade by pulling borosilicate glass tubes (inner diameter 0.94 mm, outer diameter 1.2 mm, World Precision Instruments, Sarasota, FL, USA) using a micropipette puller (Sutter Instruments, model P-2000, Novato, CA, USA). A sample volume of 5 $\mu$was typically loaded for each experiment providing a stable spray for 2 hrs. The tips were used and changed for each sample analysis. An optimum alignment (with small offset and angle of the tip located at 5-10 mm to the spectrometer inlet) is essential for spray formation and to reduce neutral droplets. A thin conducting wire of tungsten is then introduced into the sample-filled capillary to bias the sample at the desired voltage, establishing a potential difference of 1100 to 1500 V relative to the entrance-end of the inlet capillary. The inlet capillary is an 18 cm long glass transfer tube with a semiconductor inner coating (Bruker Daltonik, i.d. 0.5 mm, resistance $1000\,M \Omega$). The potential at the exit of the capillary is typically 230V lower than the one applied at the entrance of the first ion funnel. The exit of the capillary is on axis and is located at ~5 mm to the entrance ion funnel. 
The dual-ion funnels were provided by Bruker Daltonics (maXis design), together with their RF power supplies (QTIII Funnel Generator 1 and QTIII Funnel Generator 2). Homemade vacuum feedthroughs mounted on PEEK flanges were used to deliver the RF and DC signals applied to the ion optical elements. The funnels are followed by a multipole ion guide and terminated by a focusing einzel lens. The focusing lens provides a suitable beam shape for transferring the ions into the CDMS chamber. Figure~\ref{fig:other} shows the DC profile across the coaxial arrangement for transferring ions from the nanoelectrospray source to the analyzer. The velocity of the gas stream entering the vacuum chamber is estimated by measuring the velocity of charged droplets for each measurements, as they are considered unaffected by the applied voltages. The velocity of the gas stream was found in the range [360-470] m/s.\cite{doussineau2011CDMS_PEG}

\subsection{CDMS analyzer}
\subsubsection{Simulation and design of the linear charge detection array ion trap }
Initially, ion simulations were performed (SIMION simulations) \cite{dahl2000simion} in order to find the best design for the charge detection device. 

\begin{figure}
 \includegraphics[width=\columnwidth]{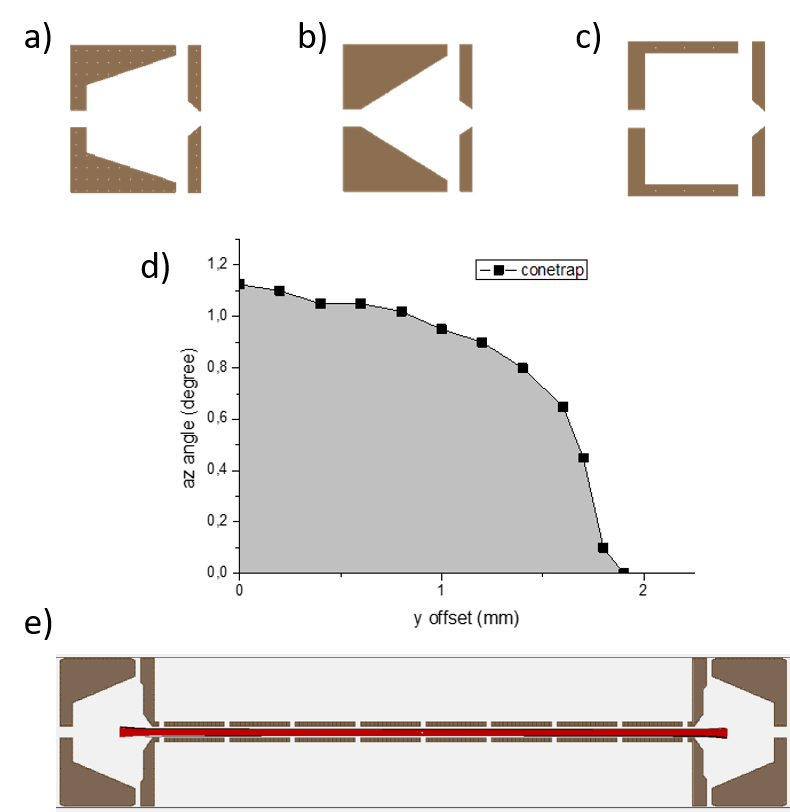}
 \caption{\label{fig:cones} a-c) Different cone geometries considered for the trapping mode. d) Acceptance of the selected cone a) according to ion beam divergence and axis offset. e) Illustration of simulated ion trajectory trapped (red) between the 2 cones.}
\end{figure}

The basic idea of the design is based on 8 tubes of 20 mm long and 4 mm inner diameter for collection of image charge of the ions in a linear array configuration. Each tube is separated by 2 mm. 4 tubes are interconnected in alternating fashion with a metal wire to carry the charge signal to the preamplifier (see Figure~\ref{fig:apparatus}). For the ion trapping, conical electrodes are placed on each side of the tubes at a distance of 3 mm. Three different geometries for trapping electrodes were tested (see Figure~\ref{fig:cones}) in terms of their capability to trap ions with deviation from on axis position (y offset) and elevation angles (az angle). The electrode geometry corresponding to Figure~\ref{fig:cones}a was found to have the best performance (Figure~\ref{fig:cones}d) and was chosen for the CDMS device. We observed that ions with an offset below 2 mm and a divergence below 1° could be trapped efficiently. For illustration, an example of a trapped ion trajectory is shown in Figure \ref{fig:cones}e (red). The typical voltages used for the 3 ions are reported in the last column of Table \ref{tab:IONsimion}.

SIMION simulations then considered 3 different typical ions produced by ESI (Table \ref{tab:IONsimion} of appendix~\ref{app:SIMION}) that covers an extended range of mass and charge. The velocity of the gas stream entering the vacuum chamber was set at ~410 m/s.\cite{doussineau2011CDMS_PEG} The terminal velocity of ions $\#1;\#2;\#3$ were estimated considering an acceleration voltages of several hundred volt in the first set of electrodes in the ion guide chamber. The mass and charges of ions $\#1;\#2;\#3$ were estimated based on previous works performed with the first generation of CDMS instrument developed in the University of Lyon.\cite{antoine2020CDMS}\\

\begin{figure}
 \includegraphics[width=\columnwidth]{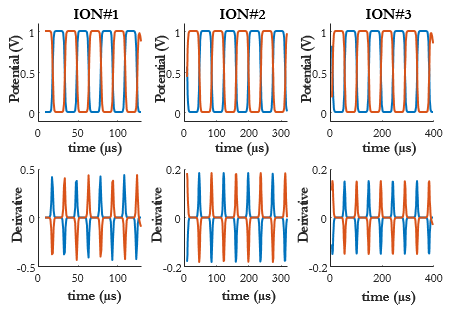}
 \caption{\label{fig:sim_trace} Simulated signal traces for the 3 ions in linear array (transmission) mode. The properties of the 3 ions can be found in Table~\ref{tab:IONsimion} of appendix~\ref{app:SIMION}. The top line corresponds to the traces out of the tubes and the bottom line corresponds to their derivatives.}
\end{figure}

First, the transmission mode was tested. Simulated signal traces were recorded for the 3 ions by measuring the potential seen by an ion flying on axis of the tubes which have alternating voltages (Vtube<< KEion) as function of the ion time-of-flight. Figure~\ref{fig:sim_trace} shows the traces with the signal for both 4-tube detectors (blue and red). We can see that typical TOFs for one tube span from 10 to 50 µs (see Table \ref{tab:IONsimion}) giving an acquisition time window for the transmission mode around 100 to 500 µs depending on the ions.

\begin{figure}
 \includegraphics[width=\columnwidth]{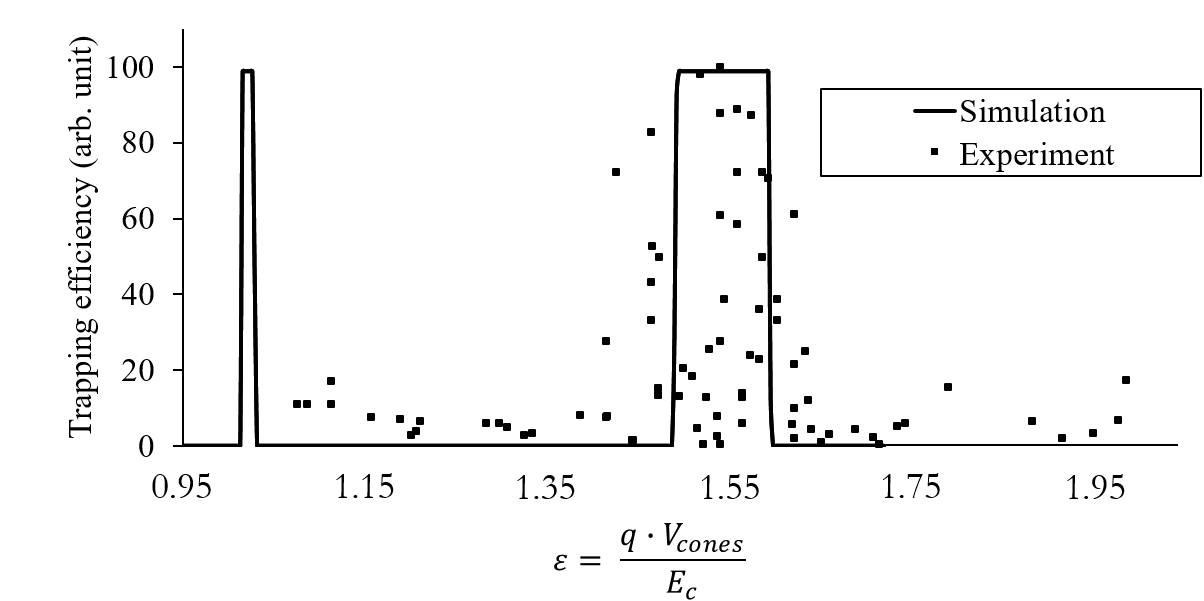}
 \caption{\label{fig:sim_trap_eff} SIMION simulation of the trapping efficiency for the linear charge detection array ion trap spectrometer as a function of the reduced parameter $\varepsilon$. Points: Experimental trapping efficiency for oscillations of single hNoVLPs ions in the trap. A trapping efficiency of 100 represents ~20 ms trapping time.
}
\end{figure}

Similarly to the previous studies with ConeTrap instruments \cite{schmidt2001conetrap,bernard2003single,REINHED201083}, we have been specifically interested in the trapping efficiency of the present linear charge detection array ion trap spectrometer using also cone electrode geometries. Figure~\ref{fig:sim_trap_eff} shows the trapping efficiency for oscillation of a single ion as a function of the potential of conic electrodes Vc (or more specifically the reduced parameter $\varepsilon$). The simulated curve of Figure~\ref{fig:sim_trap_eff} corresponds to a trapping time of 20 ms. Identically to the previous ConeTrap designs, which was operated in completely different conditions, two regions of stability are observed. A sharp regime at low cone voltages and a broader regime at higher cone voltages. These two regimes correspond to trajectories with smooth and concave mirror-like turning points. In this work, we mainly work with the broader regime at higer cone voltages (as illustrated with experimental points shown in Figure~\ref{fig:sim_trap_eff}).

\subsubsection{Implementation of  the linear charge detection array ion trap spectrometer}
This SIMION simulations led to a conceptual design for the CDMS device. Of note, the  final dimensions of the CDMS as well as the number of tubes were optimized in such way that the total capacitance of the device would fit the entrance capacitance of JFET charge sensitive transistor. With the total length of the CDMS ($\approx$ 30 cm), a total number of 8, 10, or even 12 tubes could be considered. We chose a final number of 8 tubes for simplicity of mechanical manufacturing.  Based  on the conceptual designs, SMGOP company produced the technical drawing and manufactured the entire frame including the different parts of the CDMS device. The last vacuum stage contains the CDD and is at ~$10^{-6}$ mbar. The voltages applied to the cone electrodes were adjusted until reflected trajectories were nearly parallel to incoming trajectories (see \ref{app:SIMION} for more details).
The setup was then cabled to connect the different electrodes. This connection was done with a copper wire of 1 mm diameter that connects the 4 CDD tubes to the entrance of the JFET of the preamplifier. The detectors are equipped with two preamplifiers installed in vacuum (rectangular cartridge in Figure~\ref{fig:apparatus}) and directly connected to the tubes through the copper wire. One of these preamplifiers is a commercial low-noise charge-sensitive model (Amptek CoolFET A250CF). The CDMS chamber was installed on an experimental bench and then pumped at high vacuum level ($2.5\times10^{-6}$ mbars). In order to perform first tests of the CDMS, an acquisition system using an AlazarTech acquisition card and an in-house acquisition software was used. In parallel, efforts have been made for a new high performance DAQ system from Spectroswiss \cite{Kozhinov_Nagornov_Tsybin_2025}. In addition, in-house low-noise room-temperature charge sensitive preamplifiers and derivative-amplifiers were developed and electronics for the trapping mode was also developed in-house. In the present design, 4 CDD tubes are connected to the entrance of the JFET of the cooled low-noise charge sensitive preamplifier and the 4 other CDD tubes are connected to the low-noise room-temperature charge sensitive preamplifier.

\subsection{Performance of linear charge detection array ion trap spectrometer for single viral particles analysis}

\begin{figure}
 \includegraphics[width=\columnwidth]{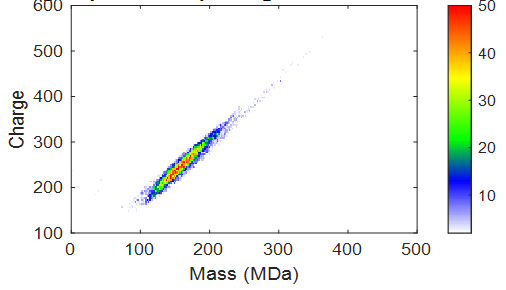}
 \caption{\label{fig:Ad(5)} 2D Mass-Charge of human Adenovirus 5 (hAdV5) ions.}
\end{figure}

We investigated two samples resembling human viruses with CDMS analysis under transmission and trapping modes, the norovirus-like particles (hNoVLPs) from the GI.1 Norwalk strain and the adenovirus 5 (hAdV5). The hNoVLPs form mainly $T=3$ assemblies of $\approx$ 10 MDa and 35-40 nm in diameter, whereas the hAdV5 form assemblies of $\approx$ 150 MDa.\cite{barnes2022adenovirus} Using the CDMS in the linear array mode, a mass distribution peaked around $\approx$ 154 MDa and charge of $\approx$ 230 e is observed for hAdV5 (see Figure~\ref{fig:Ad(5)}). With the CoolFET device, an RMS noise of 10 e (transmission mode) is observed demonstrating the high sensitivity of the device. Consequently,  the limit of detection (LoD) is around 50 charges which is up to 6-fold better than most of the single-pass measurements from the literature ($\approx$ 300 e).\cite{keifer2017CDMS_single}

\begin{figure}
 \includegraphics[width=\columnwidth]{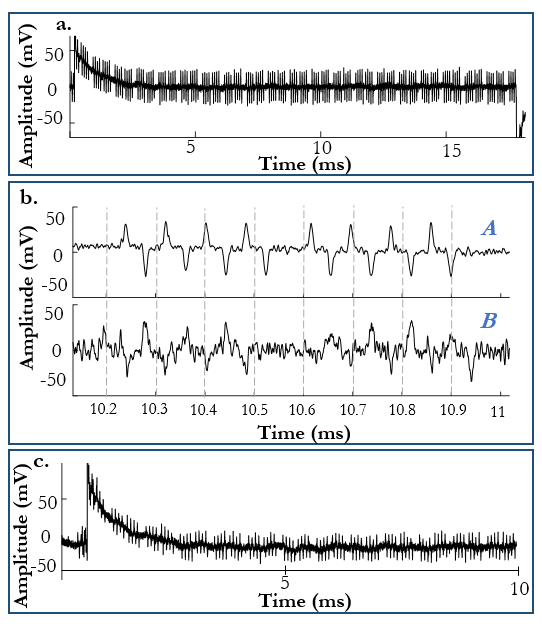}
 \caption{\label{fig:trap_Noro} a. Signal trapping trace obtained for norovirus-like particle (hNoVLP) ions (potential applied on ConeArrayTrap = 700 V). trapping in the linear array detector. b. Comparison of the resulting signal collection in A: cooled low-noise preamplifier and B: room-temperature preamplifier. c.Signal trapping trace obtained for human Adenovirus 5 ions (potential applied on ConeArrayTrap = 970 V). }
\end{figure}

To illustrate the mass measurement of a single ion that has been efficiently trapped in the linear charge detection array ion trap spectrometer, transients signals from each of the four even detector channels for a single ion obtained from hNoVLPs and hAdV5 ion is shown in Figure~\ref{fig:trap_Noro}. Each time an ion passes through the detector tubes, it induces voltage pulses corresponding to each time the ion enters or exits a single tube, with pulses on each detection channel. The hNoVLPs ion (Figure~\ref{fig:trap_Noro}.a and b) was trapped for 18 ms, during which time it traveled 20 cycles (320 single tube passes) through the trap at a velocity of 482 m/s in the field free region of the trap and had an $m/z$ value of $3.7\times 10^{5}$. The trap potential was 700 V, corresponding to a kinetic energy of 476 eV/charge. Similarly, transients signals from trapping for 10 ms of an hAdV5 ion is shown in Figure~\ref{fig:trap_Noro}.c, with a trap potential of 790 V. Both successful trapping for several ms of the two viral ions fall in the second regime of the trap potential. This result is highlighted in Figure~\ref{fig:sim_trap_eff} for experimental data of hNoVLPs ions trapping efficiency at different trap potential and compared to simulations.

\begin{figure}
 \includegraphics{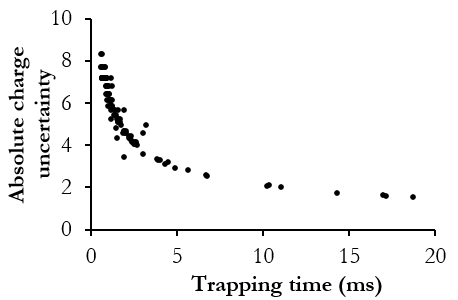}
 \caption{\label{fig:exp_trap_eff} Absolute charge uncertainty of the hNoVLPs ions according to the trapping efficiency. Trapping time of all ions with an absolute charge uncertainty less than 10 charges and as low as 2 charges for ions trapped for approximately more than 10 ms. }
\end{figure}

When we evaluate the trap performance for norovirus-like particles, it is essential that ions are trapped for an extended period of time in order to make precise mass and charge measurements. Only ions that complete at least two cycles are used for determining the performance of the ion trap, which is approximately 10\% of all ions entering the detector array in the present configuration. The uncertainty in the charge and TOF measurement decreases with an increasing number of passes through the detector tube. The distribution of trap residence times and corresponding charge uncertainties from a single detection channel for the 131 trapped ions from hNoVLPs (Figure~\ref{fig:exp_trap_eff}) illustrates the advantages of a longer trapping time to improve the precision of the charge measurement (which is the limiting parameter for CDMS performance). Based on these measurements, a charge uncertainty of 0.2 e after 1s of trapping can be extrapolated (in line with the current performance leveraged by Jarrold's group).\cite{doi:10.1021/jasms.5c00049,doi:10.1021/jasms.9b00010}

\section{Conclusion}
We described a new experimental setup to perform mass analysis of single viral particles, combining nanoelectrospray ionization to a linear charge detection array ion trap spectrometer. This ion trap spectrometer combines an eight tube detector array with conical electrodes, allowing for recording both in transmission mode (linear array) and ion trapping mode, time-of-flight (related to m/z) and charge of individual ions, thus enabling high throughput metrology of viruses at the single particle level. The devices and the geometry of the instrument have been derived from ion optics simulations. The performance of the present instrument is illustrated with hNoVLPs. \\

The linear charge detection array ion trap spectrometer, currently operating in stand-alone mode, has been successfully duplicated and is now set to be integrated with the ARIADNE-Vibe platform. This integration will leverage the high performance of the omnitrap device to prepare ions with precise alignment and optimal kinetic energy, facilitating their efficient transfer to the CDMS instrument and thereby enhancing the trapping capabilities.\cite{papanastasiou2022omnitrap}
Additionally, data from both channels will be recorded simultaneously using an upgraded high-performance data acquisition system. This new system is built upon the current DAQ technology from Spectroswiss, which is recognized as one of the most sophisticated and advanced data acquisition systems for FTMS available today.\cite{desligniere2024orbitransient} 

\begin{acknowledgments}
The research leading to these results has received funding from the European Innovation Council under the European Union Horizon Europe 2020 research and innovation program's projects \#964553 (ARIADNE-Vibe) and HORIZON‐EIC‐2022‐PATHFINDEROPEN‐01 \#101099058 (Virusong). We acknowledge Jens Hoehndorf and Manuel Chapelle from Bruker Daltonik for technical support and Serge Vialet (SMGOP) for fruitful discussions during the design of the instrument. The authors thank Jérôme Bernard for fruitful discussions regarding the operation modes of ConeTrap.
\end{acknowledgments}

\appendix
 
\section{\label{app:SIMION}Ion properties and results of the SIMION simulations}
 
 \begin{table}
 \caption{\label{tab:IONsimion}Ion properties and results of the SIMION simulations. Expected masses and charges for biomolecular ions in the range of 10 MDa to 1 GDa}
 \begin{ruledtabular}
 \begin{tabular}{l|ccc}
 Ions & ION\#1 & ION\#2 & ION\#3 \\  \hline
 Sample & DNA & Fibrils & SARS-CoV-2 \\
 Mass (MDa) & 10 & 100 & 1000 \\
 Charges (e) & 1500 & 1000 & 1000 \\
 Velocity (m/s) & 1450 & 576 & 470 \\
 KE (eV) & 109005 & 172523 & 1148680 \\
 TOF/tube (us) & 15 & 38 & 47 \\
 Cone voltage (V) & 110 & 265 & 1850 
 \end{tabular}
 \end{ruledtabular}
 \end{table}
 
\section{Viral sample production}
\subsection{Adenovirus}

HEK 293T cells were infected with human adenovirus 5 (hAdV5) containing a GFP reporter gene.  The resulting virus suspension was purified using a CsCl gradient and stored in 150 mM ammonium acetate (AmAc), 1 mM magnesium acetate pH 7.5. The purified virus was then inactivated by applying UV-C light at 254 nm with a UVPL CL-1000 crosslinker (Analytik Jena) for 40 min. Inactivation was controlled by titer assay and fluorescence microscopy for a titer ranging from 10e-2 to 10e-7. The inactivated virus was stored at -80°C.

\subsection{Norovirus capsid}

Full length VP1 of GII.17 Kawasaki, GeneBank accession number LC037415.1, was cloned and expressed using a baculovirus system as described previously \cite  {v15071482}. Baculovirus containing the recombinant VP1 gene was grown in SF9 cells and then used to infect Hi5 cells. The screted hNoVLPs were harvested and purified by subjecting the supernatant to CsCl gradient ultracentrifugation. The resulting hNoVLPs were removed, pelleted and resuspended in phosphate-buffered saline (PBS) at pH 7.5 and stored at 4°C. For measurements they were buffer exchanged into 150 mM AmAc pH 7.5.

\bibliography{aipsamp}

\end{document}